\documentstyle[12pt]{article}
\begin{document}
\font\ninerm = cmr9

\def\footnoterule{\kern-3pt \hrule width \hsize \kern2.5pt}

\pagestyle{empty}


\vskip 0.5 cm

\begin{center}
{\large\bf Quantum-gravity phenomenology with\\ gamma rays and UHE
cosmic rays\footnote{Invited lecture given at the ``International School
of Space Science: 2001 course on Astroparticle and Gamma-ray physics in space",
L'Aquila, Italy, August 30--September 7, 2001. To appear in the proceedings.}}
\end{center}
\vskip 1.5 cm
\begin{center}
{\bf Giovanni AMELINO-CAMELIA}
\end{center}
\begin{center}
{\it $^a$Dipart.~Fisica,
Univ.~Roma ``La Sapienza'',
P.le Moro 2, 00185 Roma, Italy}
\end{center}

\vspace{1cm}
\begin{center}
{\bf ABSTRACT}
\end{center}

{\leftskip=0.6in \rightskip=0.6in
In recent years several ideas for experimental searches of
effects induced by quantum properties of space-time have
been discussed. Some of these ideas concern the role in quantum
spacetime of
the ordinary Lorentz symmetry of classical flat spacetime.
Deviations from ordinary (classical)
Lorentz symmetry are now believed to be rather natural
in non-commutative space-times, models based on
String Theory and models based on Loop Quantum Gravity.
Observations of gamma rays and ultra-high-energy cosmic rays
could play a key role in the development of this
research programme.}


\newpage

\baselineskip 12pt plus .5pt minus .5pt

\pagenumbering{arabic}
\pagestyle{plain}

\section{Introduction}

Quantum-Gravity Phenomenology\cite{gacqgplect}
is an intentionally vague name for a new approach to research
on the possible non-classical (quantum) properties of spacetime.

This approach does not adopt a specific formalism for the description
of the short-distance structure of spacetime
({\it e.g.}, ``string theory", ``loop quantum gravity"
and ``noncommutative geometry" are seen as equally deserving
mathematical-physics programmes);
it is rather the proposal that quantum-gravity research should
proceed just in the familiar old-fashioned way: through small incremental
steps starting from what we know, combining mathematical-physics
studies with experimental studies to reach deeper and deeper layers
of understanding of the short-distance structure of spacetime.
For various ``historical" reasons (mostly connected with the lack
of guidance from experiments)
research on quantum gravity has wondered off this traditional
strategy: the most popular quantum-gravity approaches, such as
string theory and loop quantum gravity, could be described
as ``top-to-bottom approaches", since
they start off with some key assumption about
the structure of spacetime at the Planck scale and then they try
(with limited, vanishingly small, success)
to work their way back to ``reality",
the realm of doable experiments.
With ``quantum-gravity phenomenology" I would like to refer to all
studies that are somehow related with the ``bottom-to-top approach",
consistently with traditional strategy of physics research.

Since the problem at hand is extremely difficult (arguably the most challenging
problem ever faced by the physics community)
it appears likely that the two complementary
approaches might combine in a useful way: for the ``bottom-to-top approach"
it is important to get some guidance from the (however tentative)
indications emerging
from the ``top-to-bottom approaches", while for ``top-to-bottom approaches"
it might be very useful to be alerted by quantum-gravity phenomenologists
with respect to the type of new effects that could be most effectively
tested experimentally\footnote{It is hard for ``top-to-bottom approaches" to
obtain a complete description of low-energy physics, but perhaps it
would be possible to dig out predictions on some specific spacetime features
that appear to deserve special attention in light of the corresponding
experimental sensitivities.}.

Until very recently the idea of a quantum-gravity phenomenology,
and in particular of attempts of identification of experiments with
promising  sensitivity, was very far from the main interests of
quantum-gravity research. One isolated idea had been circulating
from the mid 1980s: it had been realized\cite{ehns,huetpesk,kostcpt}
that the sensitivity of CPT tests using the neutral-kaon
system is such that even small effects of
CPT violation originating at the Planck scale\footnote{The possibility
of Planck-scale-induced violations of the CPT symmetry has been
extensively considered in the literature.
One simple point in support of this possibility comes from
the fact that the CPT theorem, which holds in our present conventional
theories, relies on exact locality, whereas in quantum gravity
it appears plausible to assume lack of locality at Planckian scales.}
might in principle be revealed.
These pioneering works on CPT tests were for more than a decade
the only narrow context in which the implications of quantum gravity
were being discussed in relation with experiments,
but over the last 4 years several new ideas for tests
of Planck-scale physics have appeared at increasingly fast pace,
leading me to argue\cite{gacqgplect,gacQM100} that the times might be right
for a larger overall effort in this direction,
which indeed could be called ``quantum-gravity phenomenology".
At the present time (in addition to the already mentioned CPT tests)
there are several examples of experimentally accessible
contexts in which conjectured quantum-gravity effects are being
considered, including studies of in-vacuo dispersion using gamma-ray
astrophysics\cite{grbgac,billetal},
studies of laser-interferometric limits on quantum-gravity induced
distance fluctuations\cite{gacgwiFIRST,gacgwiLATEST},
studies of the role of the Planck length in the determination
of the energy-momentum-conservation threshold conditions
for certain particle-physics processes\cite{kifu,kluz,aus,gactp},
and studies of the role of the Planck length in the determination
of particle-decay amplitudes\cite{gacpion}.
These experimental/phenomenological studies might represent the cornerstones
of quantum-gravity phenomenology since they are as close as one can
get to direct tests of space-time properties, such as space-time
symmetries. Other experimental proposals that should be seen
as part of the quantum-gravity-phenomenology programme rely
on the mediation of some dynamical theory in quantum space-time;
comments on these other proposals can be found in
Refs.\cite{gacqgplect,veneziano,peri,garaytest,ahlunature,lamer}.

In these lecture notes I intend to emphasize those aspects of
quantum-gravity-phenomenology that are relevant for
the astrophysics community.
The relevant topic is the one that concerns
the faith of the Lorentz symmetry
of classical spacetime when the spacetime is quantized.
Since the Lorentz symmetry of classical flat (Minkowski)
spacetime is verified experimentally to very high accuracy,
it appears that
any deviation from classical Lorentz symmetry, which might emerge from
quantum-gravity theories, would be subject to severe
experimental constraints. As a result Lorentz-symmetry tests
are a key component of the programme
of ``quantum-gravity phenomenology"\cite{gacqgplect,rovhisto,carlip}.

My main focus here will be on the faith of Lorentz invariance
at the quantum-spacetime level. A large research effort has been
devoted to this subject. Most of these studies focus
on the possibility that Lorentz symmetry might be ``broken"
(in a sense clarified later in these notes)
at the quantum level; however, I have recently shown that
Lorentz invariance might be affected by spacetime quantization
in a softer manner: there might be no net loss of symmetries
but the structure of the Lorentz transformations
might be affected by the quantization procedure\cite{gacdsr,dsr3}.
In the following I shall describe rather pedagogically
the main differences between the broken-symmetry and
my new deformed-symmetry scenario.
In addition I will comment on the type of astrophysical
observations, involving gamma rays and ultra-high-energy cosmic rays,
which could provide evidence of such symmetry-related
quantum properties of space-time.
An exciting recent development in this area is that certain
puzzling gamma-ray and UHE cosmic-ray observations are being actively
discussed as possible first manifestations of a
quantum property of space-time.

Before going forward with these main points on my agenda for
these lecture notes, let me make a parenthetic remark, further claryfing
the objectives of quantum-gravity phenomenology:
The primary challenge of quantum-gravity phenomenology
is the one of establishing the properties of space-time at Planckian
distance scales, since most theoretical arguments suggest that
this is the characteristic scale of quantum space-time effects.
However, there is also recent discussion of the possibility that
quantum-spacetime effects might be stronger than usually expected,
{\it i.e.} with a characteristic energy scale that is much smaller
(perhaps in the TeV range!) than the Planck energy.
Examples of mechanisms leading to this possibility are found
in string-theory models with large extra dimensions\cite{led}
and in certain noncommutative-geometry models\cite{ncstrings}.
Of course, the study of the phenomenology of these models
is in the spirit of quantum-gravity phenomenology, but
it is, in a sense, to be considered as a sideline development
(and it is less challenging than the quantum-gravity-phenomenology
efforts that pertain effects originating
genuinely at the Planck scale).

\section{The faith of Lorentz symmetry in quantum spacetime}

If Nature hosts some form of ``quantization" (even just in the general
weak sense of ``non-classical" properties) of space-time,
this of course would also apply to flat spacetimes
({\it e.g.} if spacetime is in general discrete or noncommutative
then of course the particular case of flat spacetime will also
be described in the same way).
One might argue (more or less convincingly)
that quantum effects should be stronger in strong-curvature
contexts, such as the ones involving black holes,
but our capability of detailed experimental studies of such contexts
is vanishingly small. Instead, in certain flat-spacetime contexts
our experiments reach extremely high precision and therefore
even relatively small effects induced by quantum properties of spacetime
might be detectable. This is one of the key strategic points of
my view on the development
of quantum-gravity phenomenology\cite{gacqgplect,gacQM100}.

In flat quantum spacetimes a key characteristic is the role
of the Planck length, $L_p$.
If the Planck length only has the role we presently attribute
to it, which is basically the role of a coupling constant
(an appropriately rescaled version of the gravitational coupling $G$),
no problem arises for FitzGerald-Lorentz contraction,
but if we try to promote $L_p$ to the status of an intrinsic
characteristic of space-time structure (or a characteristic of
the kinematic rules that govern particle propagation in space-time)
it is nearly automatic to find conflicts
with FitzGerald-Lorentz contraction\cite{gacdsr,dsr3}.

For example, it is very hard (perhaps even impossible)
to construct discretized versions or non-commutative versions
of Minkowski space-time which enjoy ordinary
Lorentz symmetry.
Pedagogical illustrative examples of
this observation have been discussed, {\it e.g.},
in Ref.\cite{hooftlorentz} for the case of discretization
and in Refs.\cite{majrue,kpoinap}
for the case of non-commutativity.
The action of ordinary (classical) boosts
on discretization length scales (or non-commutativity length
scales) will naturally be such that
different inertial observers
would attribute different values to these lengths scales,
just as one would expect from
the mechanism of FitzGerald-Lorentz contraction.

There are also dynamical mechanisms (of the spontaneous symmetry-breaking
type) that can lead to deviations from ordinary Lorentz invariance;
it appears for example that this might be possible in String
Field Theory\cite{kosteLORENTZ}.

Both in String Theory and in Loop Quantum Gravity\footnote{As I shall argue
more carefully elsewhere\cite{gacinprep2},
in Loop Quantum Gravity there might even be a
{\underline{fundamental}}
departure from classical Lorentz invariance. This can be
deduced from studies arguing that Loop Quantum Gravity predicts
a fixed discrete spectrum of area eingevalues, independently of
the characetristic scale of curvature of the surface whose
area is being measured (and therefore also for flat surfaces in
flat spacetimes).
One of the primary implications of Lorentz invariance is that the
same experiment is seen by different observers in different
ways which are however predictably (classically) connected
by Lorentz transformations. If, for example,
a series of measurements by one observer
all give the same result of an area measurement, say the result $A_0$,
then according to classical Lorentz invariance those same
measurements should
be seen by another observer as measurements all giving
the same but different, say $A_1$, result (with $A_1$ related to $A_0$
by the appropriate boost).
When the spectrum of the area of a flat surface in a flat spacetime
is discrete this property of classical Lorentz invariance
is at risk: the results $A_0$ being all the same would reflect
the fact that one is dealing with what is
an area eigenstate for observer $O_0$, and $A_0$ should be an
eigenvalue of the area operator,
but, if the second observer $O_1$ is only minutely boosted
with respect to $O_0$, one should find that $A_1$, the boosted value
of $A_0$, could not possibly be another eigenvalue (if the boost is
small enough it will not be sufficient for reaching another
eigenvalue in the discrete list of eigenvalues that composes
the spectrum of the area operator) and it would be paradoxical
for observer $O_1$ to find systematically repeated measurement
results $A_1$.}
it is also natural to consider certain external-field backgrounds, which,
in the appropriate sense\cite{gacdsr,dsr3} (they provide a way to identify
a preferred class of inertial observers), break Lorentz invariance.

Departures from ordinary Lorentz invariance are therefore rather
plausible at the quantum-gravity level.
Here I want to emphasize that there are at least two possibilities:
(i) Lorentz invariance is broken and (ii) Lorentz invariance is deformed.

\subsection{Deformed Lorentz invariance}

In order to be specific about the differences between
deformed and broken Lorentz invariance let me focus on
the dispersion relation $E(p)$ which will naturally be modified
in either case.
Let me also assume, for the moment, that the deformation
be Planck-length induced: $E^2= m^2 + p^2 + f(p,m;L_p)$.
If the function $f$ is nonvanishing and nontrivial and
the energy-momentum transformation rules are ordinary (the ordinary
Lorentz transformations) then clearly $f$ cannot have the exact
same structure for all inertial observers. In this case one
would speak of an instance in which Lorentz invariance is broken.
If instead $f$ does have the exact
same structure for all inertial observers, then necessarily
the transformations between these observers must be deformed.
In this case one
would speak of an instance in which the Lorentz transformations
are deformed, but Lorentz invariance is preserved (in the deformed sense).

While much work has been devoted to the case in which Lorentz invariance
is actually broken, the possibility that Lorentz invariance might be
deformed was introduced only very recently by this
author\cite{gacdsr,jurekdsr,michele,dsr3,rossano}.
An example in which all details of the deformed Lorentz symmetry
have been worked out is the one in which one enforces
as an observer-independent statement the dispersion relation
\begin{equation}
L_p^{-2}\left(e^{L_p E}
+e^{- L_p E}-2\right)-\vec{p}^2e^{-L_p E}
=m^2
\label{eq:disp}
\end{equation}
In leading (low-energy) order this takes the form
\begin{equation}
E^2 - \vec{p}^2 + L_p E \vec{p}^2
=m^2~.
\label{eq:displead}
\end{equation}
The Lorentz transformations and the energy-momentum conservation
rules are accordingly modified\cite{gacdsr,dsr3,rossano}.

\subsection{Broken Lorentz invariance}

The case of broken Lorentz invariance requires fewer comments
since it is more familiar to the community.
In preparation for the following sections
it is useful to emphasize that the same dispersion relation
(\ref{eq:displead}), which was shown in Refs.\cite{gacdsr,dsr3}
to be implementable as an observer-independent dispersion relation
in a deformed-symmetry scenario, can also be considered\cite{grbgac} as
a characteristic dispersion relation of a broken-symmetry scenario.
In this broken symmetry scenario the dispersion relation (\ref{eq:displead})
would still be valid but only for one ``preferred" class of inertial
observers ({\it e.g.} the natural CMBR frame) and it would be valid
approximately in all frames not highly boosted with respect to
the preferred frame. In highly-boosted frames one might find the same
form of the dispersion relation but with different value of
the deformation scale (different from $L_p$). All this follows
from the fact that in the broken-symmetry scenario the laws
of transformation between inertial observers are unmodified.
Accordingly also energy-momentum conservation rules are unmodified.

Another scenario in which one finds broken Lorentz invariance
is the one of canonical noncommutative spacetime, in which
the dispersion relation is modified (with different deformation
term\cite{suss,luisa}), but, again, the energy-momentum
Lorentz transformation rules are not modified.
This example of noncommutative spacetime has been recently shown
to be relevant for the description of string theory in certain
external-field backgrounds (see, {\it e.g.}, Ref.\cite{ncstrings,suss}).

\section{Illustrative example: photon-pair pion decay}

Before discussing the role that observations of gamma rays
and UHE cosmic rays could play in the development
of this research area, let me clarify, in this Section,
that the differences between
the broken-symmetry and the deformed-symmetry case
can be very significant for what concerns experimental
signatures. This is also important since it proves
that the relevant astrophysics
observations might not only provide us the first manifestation
of a quantum space-time property: they might even distinguish
between different quantum pictures of spacetime.

In order to render very explicit the differences between
the broken-symmetry and the deformed-symmetry case
I consider here the simplest example in which these differences
are rather dramatic: photon-pair pion decay. I adopt in one case deformed
energy-momentum conservation\cite{dsr3}, as required by the deformed
Lorentz transformations of the deformed-symmetry case, while in the
other case I adopt ordinary energy-momentum conservation, as required
by the fact that the Lorentz transformation rules are unmodified
in the broken-symmetry case, but
for both cases I impose the same
dispersion relation (\ref{eq:displead}).

In the broken-symmetry case, combining (\ref{eq:displead})
with ordinary energy-momentum conservation rules,
one can establish a relation between
the energy $E_\pi$ of the incoming pion, the opening angle $\theta$
between the
outgoing photons and the energy
$E_\gamma$ of one of the photons
(the energy of the second photon
is of course not independent; it is given by
the difference between the energy
of the pion and the energy of the first photon):
\begin{eqnarray}
\cos(\theta) &\! = \!& {2 E_\gamma E_\gamma' - m_\pi^2
+ 3 L_p E_\pi E_\gamma E_\gamma'
\over
2 E_\gamma E_\gamma'
+ L_p E_\pi E_\gamma E_\gamma'} ~,
\label{pithresh}
\end{eqnarray}
where indeed $E_\gamma' \equiv E_\pi - E_\gamma$.
This relation shows that at high energies (starting at values of $E_\pi$
of order $(m_\pi^2/L_p)^{1/3}$) the phase space available to the decay
is anomalously reduced:
for given value of $E_\pi$ certain values of $E_\gamma$
that would normally be accessible to the decay are no longer
accessible (they would require $cos \theta > 1$).

In the deformed-symmetry case one enforces the deformed
conservation rules\cite{dsr3}
\begin{eqnarray}
E_\pi = E_\gamma + E_\gamma'~,~~~\vec{p}_\pi
= \vec{p}_\gamma + \vec{p}_{\gamma'} + L_p E_\gamma \vec{p}_{\gamma'} ~,
\label{cons}
\end{eqnarray}
which, when combined again with (\ref{eq:displead}), give rise
to the different relation
\begin{eqnarray}
\cos(\theta) &\! = \!& {2 E_\gamma E_\gamma' - m_\pi^2
+ 3 L_p E_\gamma^2 E_\gamma' + L_p E_\gamma E_\gamma'^2
\over
2 E_\gamma E_\gamma'
+ 3 L_p E_\gamma^2 E_\gamma' + L_p E_\gamma E_\gamma'^2} ~.
\label{pithreshdef}
\end{eqnarray}
Here it is easy to check that
one is never led to consider the paradoxical condition $cos \theta > 1$.
There is therefore no severe implication of the deformed-symmetry case
for the amount of phase space available for the decays
(certainly not at energies around $(m_\pi^2/L_p)^{1/3}$,
possibly at Planckian energies).

\section{An agenda for gamma-ray and UHECR studies}

The key points for the phenomenology of quantum-gravity-induced
deviations from classical Lorentz invariance are possible
deformations of the dispersion relation
and possible deformations of the energy-momentum
conservation conditions.

Whether or not there is an accompanying deformation
of energy-momentum conservation\footnote{In the case of
\underline{deformation} of Lorentz symmetry both the dispersion
relation and the energy-momentum conservation conditions are
modified simultaneously, since they both must reflect\cite{gacdsr,dsr3}
the structure
of the deformed transformation rules between inertial observers.}
a deformation of the dispersion relation is expected to
give rise to in vacuo dispersion\cite{grbgac,gacqgplect,billetal}
and, possibly (if the space-time has corresponding
structure\cite{luisa}), to birefringence.
In vacuo dispersion would provide a striking signature:
the speed of massless particles would depend on
wavelength\footnote{The ordinary dispersion relation is linear
for massless particles, and therefore $dE/dp$ is wavelength (energy)
independent. A nonlinear Planck-length-deformed dispersion
relation will instead inevitably lead to
wavelength-dependent $dE/dp$.}
and therefore
photons that we somehow know to have been emitted simultaneously
up to $\Delta_0T$ precision would reach us with relative
time delays $\Delta_1T$, where $\Delta_1T > \Delta_0T$,
and one should also find some dependence
of $\Delta_1T$ on the amount of time the photon spent
travelling in space-time ({\it i.e.} time spent under the
influence of quantum properties of space-time).
As discussed in Refs.\cite{gacqgplect,grbgac,billetal}
this type of effect can be naturally studied in the context
of observations of gamma-ray bursts and observations
of the high-energy photons emitted by certain blazars,
such a Mk421. Certain gamma-ray observatories soon to be operational
will have excellent sensitivity toward this type of effect,
and in particular GLAST\cite{glast} is planning dedicated studies.
Interest in such studies is also growing in AMS\cite{gacams}.

As discussed in the previous Section, also certain aspects
of particle-decay physics, at high energies, may carry
an important trace of quantum-space-time effects.
In that context however the implications of a dispersion-relation
deformation do depend strongly on whether there is
an associated deformation of energy-momentum conservation
({\it i.e.} depend on whether one is dealing with a scenario with
deformed symmetries or instead one is dealing with a scenario with
broken symmetries).
The outlook of these studies based on particle-decay anomalies
is described in Ref.\cite{gacpion}, also using a related data analysis
reported in Ref.\cite{dedenko}.

But perhaps the most powerful tool for the experimental investigation
of quantum-gravity-induced deviations from ordinary Lorentz invariance
is provided by ``threshold anomalies"\cite{gactp}.
It is to this topic, which deserves being discussed in detail,
that I devote the reminder of the Section.
It is intruiging to notice
that the prediction of these threshold anomalies
appears to be consistent with some puzzling results of
astrophysics observations.
In two different regimes, UHECRs and multi-TeV photons, the
universe appears to be more transparent than expected.
UHECRs should interact with the Cosmic Microwave Background
Radiation (CMBR) and produce pions. TeV photons should interact
with the Infra Red (IR) photons and produce electron-positron
pairs.  These interactions should make observations of UHECRs
with $E > 5 {\cdot} 10^{19}$eV (the GZK limit)\cite{GZK} or of
gamma-rays with $E > 10$TeV\cite{NGS} from distant sources
unlikely.  Still UHECRs above the GZK limit and
Mk501 photons with energies up to 24 TeV are observed.

A CMBR photon and
a UHE proton with $E >5 {\cdot} 10^{19}$eV should satisfy the kinematic
requirements (threshold) for pion production.  UHE protons
should therefore loose energy, due to photopion production, and
should slow down until their energy is below the GZK energy. At
higher energies  the proton's mean free path decreases rapidly
and it is down to a few Mpc at $3 {\cdot} 10^{20}$eV. Yet more
than 15 CRs have been observed with nominal energies at or above
$10^{20} {\pm} 30\%$ eV\cite{AgaWat}. There are no astrophysical
sources capable of accelerating particles to such energies within
a few tens of Mpc from us. Furthermore, if the CRs are produced
homogeneously in space and time, we would expect a break in the CR
spectrum around the GZK threshold, which is not seen.

HEGRA has detected high-energy photons with a spectrum ranging up
to 24 TeV\cite{Aharonian99} from Markarian 501 (Mk501), a BL Lac
object at a redshift of 0.034 ($\sim 157$ Mpc). This observation
indicates a second paradox of a similar nature. A high energy
photon with energy $E$ can interact with an IR background photon
with wavelength $\lambda \sim 30\mu m  (E/10TeV)$ and produce
an electron-positron pair. The mean free path of TeV photons depends
on the spectrum of the corresponding IR background.
Recent data from DIRBE\cite{Wright00,Finkbeiner,Hauser98}
and from ISOCOM~\cite{Biviano99}
suggest that the mean free path for 20TeV
photons should be much shorter than the one of 10TeV photons.
However, no apparent break is seen in the spectrum of Mk501
in the region $10$-$20$TeV.

The UHECR paradox is well established. Numerous theoretical
models, mostly requiring new physics, have been proposed for its
resolution (see Ref.\cite{Olinto} for a recent review). With
much less data, and with some uncertainty on the IR background,
the Mk501 TeV-photon paradox is less established.
However, if indeed this must be considered as a paradox,
there are no models
for its resolution, apart from the possibility that the IR
background estimates are too large. Planned experiments will soon
provide us better data on both issues. At present it appears
reasonable to assume, just as a working hypothesis,
that both paradoxes are real.

The interpretation of these paradoxes as threshold anomalies
is appealing for several reasons.
In both paradoxes low-energy photons interact
with high energy particles. The relevant reactions should take place
at a kinematic threshold. In both cases the center-of-mass
threshold energies are rather modest
and the physical processes involved are well tested and understood.
In spite of these
similarities,  so far, there is no model that explains both
paradoxes within a single theoretical scheme (unless the model
accommodates an irritatingly large number of parameters).
This appears to provide encouragement for the idea
that quantum-gravity-induced deviations from ordinary Lorentz invariance
might be responsible for both paradoxes\footnote{It is of course
also possible to consider deviations from Lorentz invariance that
do not have quantum-gravity origin\cite{Gonzalez,colgla},
but, as discussed in  Refs.\cite{gactp},
the idea of a quantum-gravity origin, besides being conceptually
appealing, leads to a natural estimate for the magnitude
of the effects, an this estimate appears to fit well
the observations (while non-quantum-gravity approaches host a large number
of parameters to be freely adjusted to obtain the needed
magnitude of the departure from Lorentz-invariance).}.

In order to illustrate the mechanism of threshold
anomalies, let us consider, for example,
the broken-symmetry case already considered in the
preceding section. I will now apply it to the kinematics
of the process of electron-positron pair production,
which is relevant for the Mk501 paradox.
Combining (\ref{eq:displead})
with ordinary energy-momentum conservation rules,
one can establish that at threshold the energy $E$ of the
hard Mk501 photon and the energy $\epsilon$ of the
soft background photon must satisfy the
relation $E \simeq m_e^2/ \epsilon + L_p m_e^6/(8 \epsilon^4)$.
The correction $L_p m_e^6/(8 \epsilon^4 )$ is indeed sufficient to
push the threshold energy upwards by a few TeV, consistently with
the observations. As shown in Refs.\cite{gactp} (and references
therein), an analogous result holds for the photopion threshold,
which is relevant for the cosmic-ray paradox.

This type of analysis provides encouragement
(of course, very preliminary)
for the hypothesis that the two paradoxes might be the first
ever manifestation of a quantum (Planck-length related)
property of spacetime.

Just like in the case of pion decay, considered in the preceding
Section, also for the evaluation of threshold anomalies
there are large quantitative differences (which will be discussed
in detail in a paper now in preparation\cite{gacinprep})
between the case in which Lorentz symmetry is broken and
the case in which Lorentz symmetry is deformed.
More accurate information on the paradoxes, such as the one that
will be provided by Auger\cite{auger},
can therefore even start pointing
us toward the proper language for the description
of the short-distance (quantum) structure of spacetime.

Experimental studies such as the ones planned by Auger
will also in general clarify whether the origin
of the paradoxes is indeed kinematical. I want to stress that,
in this respect,
it is important to get high-quality data in the
neighbourhood of the expected GZK cutoff,
perhaps even more than establishing how far (how high in
energy) the cosmic-ray flux extends.
In fact, the kinematical mechanism of threshold anomalies
leads to the definite general prediction that nothing at all
particular should happen at the GZK scale,
since
the GZK threshold is simply moved forward (or eliminated
all together\cite{gactp,ngthresh})
by the deviations from classical Lorentz invariance.
Other attempts of explaining the cosmic-ray paradox
instead must coexist with the GZK threshold and therefore
(unless huge parametric fine-tuning is allowed)
will inevitably predict at least some peculiarity
to occur at the GZK scale.

\end{document}